\documentclass[fleqn,twoside]{article}
\usepackage{espcrc2,epsfig}

\usepackage{graphicx}
\usepackage[figuresright]{rotating}

\def\fm{{\rm fm}}
 
\def\csw{c_{\rm sw}}

\newcommand{\be}{\begin{equation}}
\newcommand{\ee}{\end{equation}}
\newcommand{\bea}{\begin{eqnarray}}
\newcommand{\eea}{\end{eqnarray}}

\newcommand{\eq}[1]{eq.\,(\ref{#1})}

\newcommand{\figurebox}[2]{\fbox{\vbox to#2in{\hbox to #1in{\hfil}\vfil}}}
\newcommand{\lesssim}{\raisebox{-.6ex}{$\stackrel{\textstyle{<}}{\sim}$}}

\newcommand{\Nf}{N_{\rm f}}

\newcommand{\mps}{m_{\rm P}}

\newcommand{\mq}{m_{\rm q}}
\newcommand{\mst}{m_{\rm s}}

\newcommand{\gbarSF}{\overline{g}_{\rm SF}}

\newcommand{\AmS}{{\protect\the\textfont2
  A\kern-.1667em\lower.5ex\hbox{M}\kern-.125emS}}

\title{
\vspace{-3.1cm}
\begin{flushright}
       {\normalsize LTH~535}    \\[-0.2cm]
\end{flushright}
       \vspace{0.7cm}
       Cost of dynamical quark simulations: O($a$) improved Wilson
       fermions\thanks{Contribution to panel discussion at ``Lattice
       2001'', Berlin}}

\author{H. Wittig\address{Division of Theoretical Physics, Department
        of Mathematical Sciences, University of Liverpool, UK}
       (for the UKQCD, QCDSF and ALPHA Collaborations)
}       
\begin{document}

\begin{abstract}
I report on cost estimates and algorithmic performance in simulations
using 2~flavours of non-perturbatively O($a$) improved Wilson quarks
together with the Wilson plaquette action.

%\vspace{1pc}
%\vspace{-.4cm}
\end{abstract}

% typeset front matter (including abstract)
\maketitle

%\section{INTRODUCTION}

Several collaborations have employed O($a$) improved Wilson fermions,
following the non-perturbative determination of the clover coefficient
$\csw$ for $\Nf=2$ flavours in the range $\beta\ge5.2$
\cite{impr:csw_nf2}. UKQCD, QCDSF and JLQCD perform simulations on
physically large volumes ($L\;\lesssim\;1.5\,\fm$) using periodic
boundary
conditions~\cite{dspect:ukqcd_csw202,qcdsf_lat00,jlqcd_lat00}, whereas
ALPHA use Schr\"odinger functional (SF) boundary conditions on small
volumes ($L\ll1\,\fm$) \cite{alpha_comp,alpha_Nf2_lett}. In the
following I shall discuss the two types of simulations separately.
%
%\vspace{-0.1cm}
%
\section{RESULTS FROM UKQCD \& QCDSF}

Run parameters for the UKQCD and QCDSF simulations are listed in
Table~I of \cite{dspect:ukqcd_csw202} and Table~2 of
\cite{qcdsf_lat00}, respectively. Additional runs, which are included
here, have since been performed, and their details will be published
elsewhere.
%The parameters for some of UKQCD's runs were chosen such as
%to keep the lattice spacing in units of $r_0$ fixed
%\cite{dspect:ukqcd_csw202}. This makes it possible to study the cost
%of the simulations as a function of the pion mass for constants
%physical volume. 

Integrated autocorrelation times, $\tau^{\rm int}$, for hadron masses,
are poorly known. One therefore relies on $\tau^{\rm int}$ estimated
from the average plaquette, which shows an unexpected slight {\it
decrease} for smaller quark masses (see Table~II of
\cite{dspect:ukqcd_csw202}). Given the poor understanding of
autocorrelation times, UKQCD have chosen a constant separation of 40
HMC trajectories between ``independent'' configurations for all data
sets. Thus, any scaling of $\tau^{\rm int}$ with the quark mass (or,
equivalently, $a\mps$) has {\it not} been folded into the cost
analysis.

The number of operations per independent configuration is modelled
according to
\be
  \frac{N_{\rm ops}}{\hbox{ind. cfg.}} = C\left(
  \frac{L}{a}\right)^{z_1}\, \left( \frac{1}{a\mps}\right)^{z_2}.
\label{eq_costformula}
\ee
The available run data on $L/a=24$ provided by QCDSF are as yet not
sufficient to constrain the $L/a$ dependence well enough, and hence we
have used $L/a=16$ only, setting the coefficient $z_1$ equal to the
value quoted in ref.~\cite{SESAM_cost}, i.e. $z_1=4.55$. The
performance observed by QCDSF on $L/a=24$ is consistent with this
value.

\begin{figure}[tb]
\vspace{-3.8cm}
\hspace{0.5cm}
\centerline{
\psfig{file=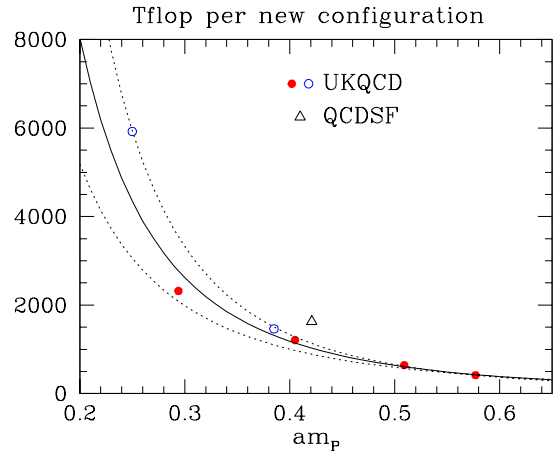,width=10.3cm}
}
\vspace{-2.0cm}
\caption{Fits to the cost formula using UKQCD's run data
\protect\cite{dspect:ukqcd_csw202} (full circles, solid line). The
dotted lines represent the uncertainties estimated by including the
open circles.
\label{fig_ukqcdfit}}
\vspace{-0.7cm}
\end{figure}

Fits to the prefactor~$C$ and~$z_2$ in \eq{eq_costformula} were
performed to the UKQCD subset of the data. Update times were converted
into $N_{\rm ops}$ using the CRAY T3E's sustained speed of 275
Mflops per processor (32bit and assembler). This yields
\be
   C=0.31(7)\,\hbox{Gflop},\quad z_2=2.77(40),
\label{eq_ukqcdfit}
\ee
where the errors have been estimated by fitting different subsets of
data points. The corresponding curves are shown in
Fig.~\ref{fig_ukqcdfit}. The estimate for $z_2$ agrees with the value
quoted in \cite{SESAM_cost}, whereas the prefactor~$C$ in
\eq{eq_ukqcdfit} is roughly 10 times larger. The reason for this is so
far unknown.

In order to estimate the CPU effort required to repeat the quenched
benchmark for the light hadron spectrum~\cite{qspect:CPPACS}, we
assume that the smallest lattice spacing and quark mass each account
for 50\% of the total. Furthermore, using O($a$) improvement implies
that one can use larger lattice spacings without compromising the
continuum extrapolations. The following estimates are based on 400
configurations on a smallest lattice spacing of $a=0.07\,\fm$, with
$L/a=48$, and a minimum pion mass of $am_{\rm P}^{\rm min}$,
corresponding to a dynamical quark mass $\mq$:
\begin{center}
\begin{tabular}{ccc}
\hline
$am_{\rm P}^{\rm min}$ & [Tflops\,years] & $\mq$ \\
\hline
0.17  & $\phantom{5}95$ & $\mst/2$ \\
0.12  &            250  & $\mst/4$ \\
0.09  &            550  & $\mst/8$ \\
\hline
\end{tabular}
\end{center}
%
%The total CPU effort is thus of O(100) Tflops\,years for quark masses
%$\mq\lesssim\mst/4$.

\section{RESULTS FROM ALPHA}

ALPHA simulate massless quarks on small volumes with SF boundary
conditions. The box size~$L$ is thus the only scale in the problem. In
particular, the condition number of the fermion matrix is determined
by~$L$, which implies that the r\^ole of $1/a\mps$ in
\eq{eq_costformula} is taken over by $L/a$. The appropriate cost
formula for the SF is therefore
\be
%  \frac{N_{\rm ops}}{\hbox{ind. cfg.}} = C^\prime\left(
%  \frac{L}{a}\right)^{z}.
  {N_{\rm ops}}/{\hbox{ind.\,cfg.}} =
   C^\prime\left({L}/{a}\right)^{z}.
\label{eq_costSF}
\ee
A detailed algorithmic study, including a cost analysis, has been
published in \cite{alpha_comp}. ALPHA were able to extract precise
autocorrelation data for the relevant observable, i.e. the running
coupling in the SF scheme $\gbarSF(L)$. For $\gbarSF^2\approx1$ one
finds that $\tau^{\rm int}\approx2$ trajectories with a relative error
of 5--10\%.

\begin{figure}[tb]
\vspace{-3.8cm}
\hspace{0.5cm}
\centerline{
\psfig{file=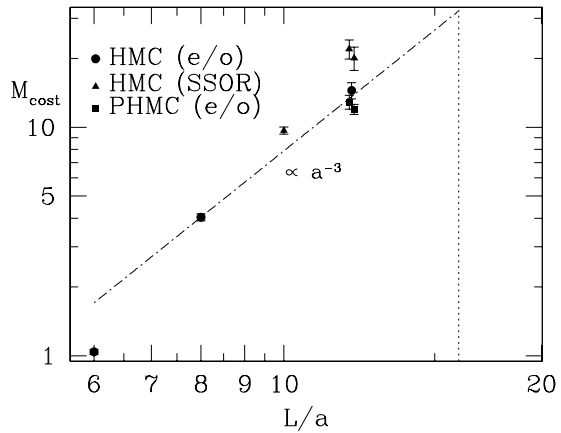,width=10.3cm}
}
\vspace{-2.0cm}
\caption{Cost versus $L/a$ from ref.~\protect\cite{alpha_comp}.
\label{fig_Mcost}}
\vspace{-0.7cm}
\end{figure}

ALPHA have used an alternative measure of the cost of their
simulations. The quantity $M_{\rm cost}$ defined in eq.\,(3.1) of
\cite{alpha_comp} is expected to differ from \eq{eq_costSF} by an
overall factor $(L/a)^3$. Fig.\,\ref{fig_Mcost}, taken from
\cite{alpha_comp}, shows a plot of $M_{\rm cost}$ versus $L/a$ for
$\gbarSF^2(L)\approx1$. It suggests a scaling of $M_{\rm{cost}}
\propto (L/a)^3$ (dashed line in Fig.\,\ref{fig_Mcost}), which implies
$z\approx6$ in \eq{eq_costSF}.

The benchmark for ALPHA is the determination of the running of
$\alpha_s$ and the extraction of the $\Lambda$-parameter. The results
of \cite{alpha_Nf2_lett} imply that lattice sizes of $L/a=16-20$
should be sufficient to determine the step scaling function $\sigma$
for $\Nf=2$ flavours with similar accuracy as in \cite{mbar:pap1}. The
total CPU effort is estimated to be of the order of 0.1 Tflops\,years,
which is within reach on machines like APE1000. This estimate does
not, however, include the computation of a low-energy scale such as
$f_\pi$ for $\Nf=2$, which is necessary to express $\Lambda$ in
physical units.

\smallskip\par\noindent
{\bf Acknowledgements} I thank Alan Irving, Karl Jansen, Dirk Pleiter
and Rainer Sommer for their help in preparing this contribution.

\end{document}